\documentclass[prl,twocolumn,floatfix,showpacs,amsmath,amssymb] {revtex4}
\usepackage{graphicx}

\begin{document}

\title{Equilibrium glassy phase in a polydisperse hard sphere system}
\author{Pinaki Chaudhuri, Smarajit Karmakar, 
Chandan Dasgupta, H.R. Krishnamurthy, A.K. Sood}
\affiliation{Centre for Condensed Matter Theory,
Department of Physics, Indian Institute of Science, Bangalore 560012
India}

\begin{abstract}

The phase diagram of a polydisperse hard sphere system is examined
by numerical minimization of a discretized form of
the Ramakrishnan-Yussouff free energy functional.
Crystalline and glassy local minima of the
free energy are located and the
phase diagram in the density--polydispersity plane is mapped out
by comparing the free energies of different local minima.
The crystalline phase disappears and the glass becomes
the equilibrium phase beyond
a ``terminal'' value of the polydispersity.
A crystal to glass transition is
also observed as the density is increased at high polydispersity.
The phase diagram obtained in our study
is qualitatively similar to that of hard spheres in a
quenched random potential.

\end{abstract}


\maketitle

Colloidal suspensions of polystyrene spheres coated with thin 
polymeric layers effectively 
behave as hard sphere systems\cite{hard}, 
exhibiting fluid and crystalline phases in equilibrium. 
For such systems, glass (amorphous solid) is believed to be a 
metastable phase\cite{megen}, resulting from structural arrest.
However, in recent theoretical
studies\cite{fab,cdgq} of hard spheres in the presence of a 
quenched random potential, an equilibrium
glassy phase was observed at high disorder strengths. 
While this phenomenon has not yet been
confirmed experimentally, it brings forth the 
question of whether even the presence of annealed
disorder in a system of hard spheres can 
result in an equilibrium glassy phase. 
The annealed disorder can be realized if the hard spheres 
have different sizes, which is the case for
most colloidal suspensions. The size dispersity can be modeled
by assuming that the diameters ($\sigma$) are sampled from a 
continuous distribution
$p(\sigma)$, characterized by a parameter $\delta$, known as 
the polydispersity, and defined as the ratio of the standard deviation 
and the mean of the diameter distribution.
In this paper we present calculations that show that for a system 
of polydisperse hard spheres, 
one does obtain an equilibrium glassy
phase. In fact, our results suggest that for hard spheres, 
quenched and annealed disorder lead to 
qualitatively similar phase behavior.

The equilibrium phase behavior of polydisperse hard spheres have generated a 
lot of interest in recent times. Experiments have shown that for such a system,
crystallization does not occur beyond a {\it terminal} polydispersity of 
$\delta_t\approx0.12$\cite{pusey}. Particle simulations\cite{phan} and 
density functional studies\cite{mcrae}
indicate a similar behavior, except that there is no consensus on the
value of $\delta_t$. 

It has been observed numerically that the typical height of nucleation barriers 
for the formation of polydisperse crystals grows anomalously as the 
polydispersity increases, 
thus suppressing crystallization at high polydispersities\cite{auer}.
This feature has been utilized in experiments\cite{weitz,vanb} 
by using polydisperse hard spheres
as a suitable system for studying the glass transition. 
Molecular dynamics simulations\cite{sear} 
of a supercooled system of polydisperse hard spheres also 
suggest the presence of dynamical heterogeneity, believed to be
a signature of glassiness.

Recently, free energy calculations\cite{barwar} found the occurrence of 
re-entrant melting (transition from crystal to liquid as the density is
increased) near $\delta=\delta_t$. This conclusion is opposed by
calculations that suggest that the equilibrium phase at high polydispersity 
corresponds to a {\it fractionated} state\cite{fasolo}.
Till now, numerical evidence for fractionation has been found only
in simulations\cite{kofke} in the grand canonical ensemble, 
which is different from the situation in typical experiments.
In a recent experiment on polydisperse hard-sphere colloids 
in a confined geometry\cite{kegel}, re-entrant 
melting was observed at the colloidal monolayer adjacent to the surface, 
although no crystallization occurred 
in the bulk. It was claimed that such a phenomena could be observed in the 
bulk as well, only if one 
waited for sufficiently long time to allow crystallization to occur. 
The presence of the wall 
resulted in lowering of the 
nucleation barrier, thereby allowing crystallization 
and re-entrant melting to occur.

Although there have been suggestions\cite{mcrae,barwar} that at high densities 
or at high polydispersities,
the equilibrium phase is a glass,  
there have been no calculations to substantiate this possibility. We have
addressed this question within the framework of density
functional theory (DFT), using the Ramakrishnan-Yussouff (RY) 
free energy functional\cite{rama}. 
In DFT, the free energy is expressed 
as a functional of the time-averaged 
local density $\rho({\bf r})$.
The RY free energy functional is given by :
\begin{eqnarray}
\beta F &=& \int{d {\bf r}\{\rho({\bf r})
\ln (\rho({\bf r})/\rho_0)-\delta\rho({\bf r})\} }  \nonumber \\
&-&\frac{1}{2}\int{d {\bf r} \int {d{\bf r}^\prime
C({|\bf r}-{\bf r^\prime|}) \delta \rho ({\bf r}) \delta
\rho({\bf r}^\prime)}} . \label{rydef}
\end{eqnarray}
Here, we have defined $\delta \rho ({\bf r})\equiv \rho({\bf r})-\rho_0$ as the
deviation of  ${\rho(\bf r})$ from $\rho_0$,
the density of the uniform liquid, and
taken the zero of the free energy at its uniform liquid value.
In Eq.{\ref{rydef}), $\beta=1/(k_B T)$,
$T$ is the temperature and $C(r)$
is the direct pair correlation
function of the uniform liquid at density $\rho_0$.

In the polydisperse limit, the RY functional becomes\cite{mcrae}
\begin{eqnarray}
\beta F &=& \int{d {\bf r}}\{\rho({\bf r})
\ln (\rho({\bf r})/\rho_0)-\delta\rho({\bf r})\}   \nonumber \\
&-&\frac{1}{2}\int{d {\bf r}} \int {d{\bf r}^\prime}  
\overline{C({|\bf r}-{\bf r^\prime|})} \delta\rho({\bf r}) \delta\rho({\bf r}^\prime)
\end{eqnarray}
where
$\overline{C(r)}=\int{d{\sigma_i}}\int{d{\sigma_j}}p({\sigma_i})p({\sigma_j})
C_{ij}(r)$,
$C_{ij}({|\bf r}-{\bf r^\prime|})$ being the partial direct correlation
function between spheres of size $\sigma_i$ and $\sigma_j$.

In order to carry out numerical work, we discretize our system. We
introduce for this purpose a simple cubic computational mesh of size
$L^3$ with periodic boundary conditions.
On the sites of this mesh, we define density variables
$\rho_i \equiv \rho({\bf r}_i) h^3$, where $\rho({\bf r}_i)$ is the
density at site $i$ and $h$ the spacing of the
computational mesh. In terms of these quantities, the discretized 
form for the RY functional takes the form
\begin{eqnarray}
\beta F &=& \sum_i \{\rho_i \ln (\rho_i/\rho_L) -
(\rho_i-\rho_L)\} \nonumber \\
&-&\frac{1}{2} \sum_i \sum_j {\overline{C_{ij}}}(\rho_i-\rho_L)(\rho_j-\rho_L),
\label{discr}
\end{eqnarray}
where the sums are over all the sites of the computational mesh,
$\rho_L \equiv \rho_0 h^3$, 
and $\overline{C_{ij}}$ is the discretized form of
the direct pair correlation function $\overline{C(r)}$ of the uniform liquid.
Our algorithm\cite{cdgx} for the minimization of the discretized 
RY functional converges to the local free-energy minimum whose
basin of attraction 
contains the initial state.
It is known that, for the monodisperse hard sphere system, 
numerical minimization of the discretized form of the RY functional
yields a crystallization transition at dimensionless density 
$\rho_l \equiv \rho_0 \sigma^3 
\simeq 0.945$ if the 
mesh size $h$ is sufficiently small\cite{pinski,cdgx}. Dasgupta
and Valls\cite{cdgg} also obtained glassy minima for monodisperse
hard sphere systems using this minimization scheme.

We have extended this method of calculation to  
a system of polydisperse hard spheres.
In our calculation, the particle size distribution is chosen to be of 
the Schultz type :
$p(\sigma)={\gamma}^{\alpha}{\sigma}^{\alpha-1}
{\exp(-\gamma\sigma)}/{\Gamma(\alpha)}$,
where $\alpha=1/\delta^2$ and $\gamma=\alpha/\bar{\sigma}$, 
$\delta$ being the polydispersity
and $\bar{\sigma}$, the mean diameter of the system of particles. 
For the average direct correlation function 
$\overline{C(r)}$ for a polydisperse hard sphere system, we
use the analytical expression derived using the
Percus-Yevick approximation\cite{blum}.

To study the stability of the crystalline minimum, 
we use the fcc structure as an input
for the free-energy minimization. The fluid-to-crystal transition occurs when
the free energy of the minimum obtained this way
becomes negative (i.e. lower than that
of the uniform fluid).
For each value of $\delta$, the dimensionless density $\rho_l$ at which this 
happens is identified.
The resultant phase diagram is shown in Figure~\ref{fig1}.
In another method of calculating the free energy\cite{mcrae} of the crystal, 
the local density $\rho({\bf r})$
is approximated as a sum of Gaussian profiles: 
$\rho({\bf r})={A}/{\pi^{3/2}\epsilon^{3}}\sum_{R}
\exp[-{({\bf{r}}-{\bf R})^{2}}/{\epsilon^{2}}]$. In this expression,
$A=(1+\eta)\rho_0 v_0$, where $\eta$ is the density change at freezing and
$v_0$ is the unit cell volume of a fcc lattice with spacing $a$, 
$\epsilon$ is the
width of the Gaussians and $\{\bf{R}\}$ are the lattice points.
The RY free energy functional is then minimized with respect 
to the parameters $\epsilon$, $\eta$, and $a$
to find the values of $\{\rho_l,\delta\}$ where the free energy of the 
crystal becomes negative. In this case, 
$\rho_l \equiv {\rho_0}<\sigma^3>$ is the polydisperse liquid
density, $<\sigma^3>$ being the
third moment of the distribution $p({\sigma})$.
As shown in Fig.\ref{fig1}, the results from the grid-minimization and
the Gaussian approximation agree quite well, thereby establishing the
correctness of the results obtained by the grid method. 
Our results corroborate  
earlier density functional calculations\cite{mcrae} - 
there is no crystallization
beyond a terminal polydispersity $\delta_t \simeq 0.048$. 
Moreover, our calculations, using both
grid-minimization and Gaussian approximation, clearly indicate that for 
values of $\delta$ slightly lower than $\delta_t$,
there is re-entrant melting at high densities, confirming the result
of an earlier free-energy calculation\cite{barwar}.
\begin{figure}[htbp]
\vspace*{-0.4cm}
\includegraphics[height=7cm,width=5.5cm,angle=-90]{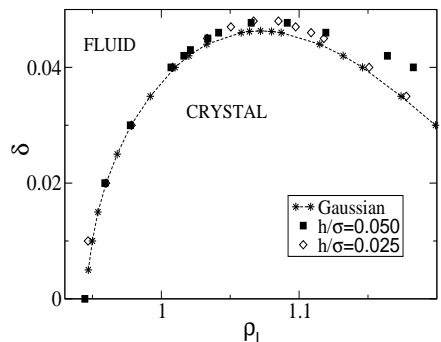} \\
\vspace*{-0.5cm}
\caption{Fluid-to-crystal transition points in the 
dimensionless density ($\rho_l$)--polydispersity ($\delta$) 
plane. Data for two values of the mesh-spacing $h$ used in 
the grid minimization, and the results obtained from the
Gaussian approximation are shown. In all the three
cases, re-entrant melting is observed at high density 
and high polydispersity.}
\label{fig1}
\end{figure}

In the density functional calculations, the value of
the terminal polydispersity comes out to be much lower than what 
is seen in experiments. A possible
reason for this difference is 
that the average direct correlation function $\overline{C(r)}$ used in 
our calculations underestimates inter-particle correlations for relatively
large values of $\delta$. The use of a
more accurate $\overline{C(r)}$ would probably 
shift the terminal polydispersity to a higher value 
without changing other features of the phase diagram.

The question that now needs to be answered is: What is the equilibrium 
phase of the polydisperse
hard sphere system at high densities, beyond the point of 
re-entrant melting, and also
at polydispersities where crystallization does not occur? 
Does it remain liquid or does it become a glass?
To answer this question, one needs to locate glassy minima of the free
energy and compare their free energy  with that of
the uniform liquid.

In the density functional framework, the glassy state corresponds to
local minima with inhomogeneous density distributions that exhibit 
strong non-periodicity~\cite{cdgx,cdgg}.
To locate such minima, one first needs to construct 
appropriate density configurations $\{\rho_i\}$ to be used as
inputs for the minimization.
In a recent density-functional study of the
glass transition in monodisperse hard sphere systems\cite{kim}, 
particle configurations from molecular dynamics (MD)
simulation were used to produce the density field 
for calculating the free energy using
the Gaussian approximation for the local density.
Similarly, in our discretized method for doing the DFT calculation,
the input density field was constructed from MD
simulation of polydisperse hard spheres.
The simulations
were carried out for $470$ particles with their diameters sampled
from the Schultz distribution with $\delta=0.0289$.
Using a modified form of the 
Stillinger-Lubachevsky compression algorithm \cite{still},
highly dense configurations of polydisperse hard spheres could be created. 
The input density field $\{\rho_i\}$ was calculated from these configurations
using a computational grid of mesh-spacing $h \approx 0.05{\bar{\sigma}}$. 
The RY functional
was then minimized in the space of the resulting 
$4.096\times10^6$ density variables to
obtain the density configuration at the local minimum of the free energy.
As in the case
of monodisperse hard spheres\cite{us}, we obtain, in this case also, 
free-energy minima
with the $\{\rho_i\}$ having a glassy structure.
The structure of a local minimum may be characterized by the 
two-point 
correlation function $g(r)$ of the  local density variables 
$\{\rho_i\}$ at the minimum. This function 
is defined as $g(r)=<\rho(0)\rho(r)/{\rho_0}^2>$. 
In Fig.~\ref{gglass}, we have plotted the $g(r)$ for a glassy free energy
minimum obtained for $\delta=0.0289$. 
The glassy nature of the density distribution 
is indicated by the split second peak of the $g(r)$, 
with the sub-peaks occurring at $1.77\bar{\sigma}$ 
and $2.02\bar{\sigma}$. These numbers are close to those for a 
dense random packing of monodisperse hard 
spheres, where the sub-peaks occur at 
around $1.7\sigma$ and $2\sigma$\cite{john}. 

\begin{figure}[htbp]
\vspace*{-0.4cm}
\includegraphics[height=6cm,width=5.2cm,angle=-90]{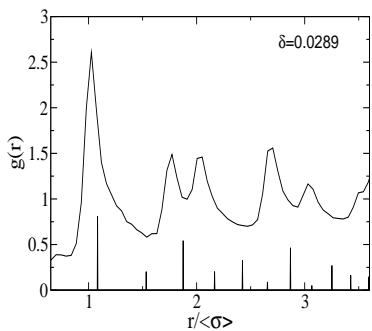} \\
\vspace*{-0.2cm}
\caption{The two point density correlation 
function $g(r)$ for a glassy minimum obtained
for a polydispersity value of $\delta=0.0289$ and dimensionless 
liquid density $\rho_l=1.11$.  
Also plotted in the figure are the peaks of the pair distribution function
(not drawn in the same scale as $g(r)$ for the glass)
for a fcc lattice at this density.}
\label{gglass}
\vspace*{-0.3cm}
\end{figure}

The density configuration $\{\rho_{i}\}$ at the local minimum for
$\delta=0.0289$ was thereafter used as input to search for similar 
glassy minima 
at other values of the polydispersity. 
The density at which the free energy of the glassy structure becomes lower 
than that of the uniform liquid defines
the point of the liquid-to-glass transition. Fig.~\ref{fgp} shows a plot of
the liquid-to-glass transition density for various values of $\delta$.
As shown in the plot, one obtains glassy minima 
for polydispersities higher than $\delta_t$, the terminal value beyond which 
crystallization does not occur. 
Also the glass transition point shifts to 
higher densities as a function of increasing polydispersity and at 
large enough polydispersity, near $\delta=0.10$, the glass transition point
approaches the random close packing limit for polydisperse 
hard spheres\cite{phan,santiso}.
When the density configuration for the minimum for $\delta=0.02$ is used as  
input for minimizing the RY functional for $\delta=0.0$, i.e.
for a system of monodisperse hard spheres, we obtain a 
free energy minimum whose $\{\rho_i\}$ correspond
to a polycrystalline structure.

\begin{figure}[htbp] 
\vspace*{-0.4cm}
\includegraphics[height=6cm,width=5.5cm,angle=-90]{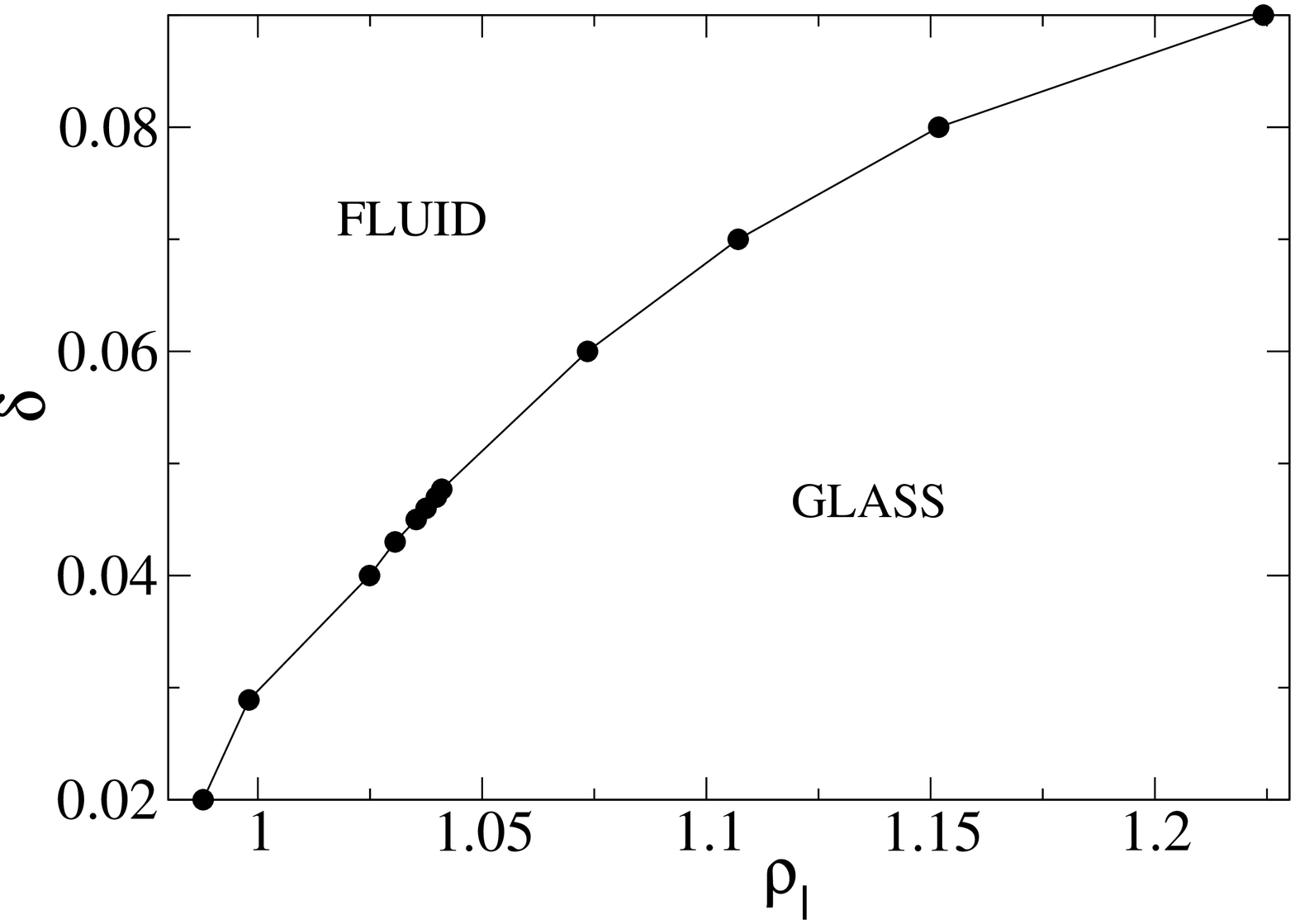} \\
\vspace*{-0.3cm}
\caption{Glass transition points, plotted in the dimensionless 
density ($\rho_l$)--polydispersity ($\delta$)
plane. At these points, the RY free energy of the glassy minimum
becomes negative. 
The line drawn through the data points is a guide to the eye.}
\label{fgp}
\end{figure}

Having determined the densities ($\rho_l$)
beyond which the crystalline
and glassy phases, respectively, have lower free energies compared
to that of the homogeneous liquid phase, the
next task is to determine the final phase diagram, i.e., to determine
which phase (crystal, glass or liquid)
is the thermodynamically stable one at different points
in the $(\delta, \rho_l)$ plane.
At any density, if more than one local minima
of the free energy are present, the thermodynamically stable phase corresponds
to the one with the lowest free energy. Using this criterion, we have
obtained the full phase diagram of the polydisperse hard sphere
system, which is shown in Fig.~\ref{final}. The inset in 
Fig.~\ref{final} shows, for example, how the crystal-to-glass transition 
line is located -- for each polydispersity, the free energies of the
glassy and crystalline minima at different values of $\rho_l$ are compared and
the value of $\rho_l$ at which the free energy of the glassy minimum 
becomes lower than that 
of the crystalline minimum defines the crystal-to-glass transition point.

\begin{figure}[htbp] 
\vspace*{-0.45cm}
\includegraphics[height=8cm,width=6cm,angle=-90]{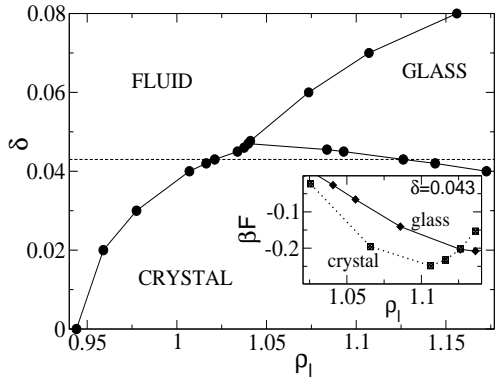} \\
\vspace*{-0.3cm}
\caption{The overall phase diagram of a system of polydisperse
hard spheres in the dimensionless density ($\rho_l$)--polydispersity ($\delta$)
plane showing the presence of the three phases: uniform liquid, crystal,
and glass.
The lines drawn to denote the phase boundaries are only meant to be guides
to the eye. 
The inset shows the free energies of the
the glassy and crystalline minima for $\delta=0.043$ (shown by the dotted
line in the main figure). The crystal is the equilibrium phase
till $\rho_l \simeq 1.126$, after which the 
glass becomes the equilibrium phase. 
}
\label{final}
\vspace*{-0.4cm}
\end{figure}

It is important to note that the re-entrant melting shown in Fig.~\ref{fig1}
is not present in the final phase diagram. Now, as one moves
along a line of fixed polydispersity (for example, $\delta=0.043$
as shown  by the dotted line in Fig.~\ref{final}),
one first encounters the fluid phase, then a crystalline phase 
and at even higher densities, glass becomes the stable phase 
(as is clear from
Fig.~\ref{fgp}). 
The other significant feature of the phase
diagram is that,
at a fixed high dimensionless density ($\rho_l > 1.05$),
if we increase the polydispersity $\delta$, the crystal undergoes a transition
to the glassy state.
We find that the crystal--glass transition line is below the
fluid--crystal re-entrant line in Fig.\ref{fig1}.
A possible reason for 
this behavior is that as the polydispersity increases, the number of 
smaller particles in the system increases and their presence makes
the glassy phase to be entropically favorable as compared
to  the crystalline phase.
Very similar features, including the presence
of an equilibrium glassy phase and crystal-to-glass transitions upon
increasing density or disorder, have been found\cite{cdgq,fab} in the
phase diagram of a system of hard spheres in a quenched random potential.

Our calculations do not take into account the possibility of 
having fractionated phases. 
We need to compare the free energy of the 
glassy phase with that of fractionated phases 
at high polydispersities to determine which one is the
equilibrium phase. 
A proper procedure for doing such calculations in the framework of 
DFT is not yet available.
However, one must remember that for many glass-forming liquids, 
phase separation is 
possible but may not be observed in experimental time scales.
This may be the case for this system as well.

In summary, we have shown, using density functional theory, 
that re-entrant melting occurs at high densities 
in a system of polydisperse hard spheres if only the crystal and liquid states
are considered. However, the re-entrant liquid state is replaced by a glassy
state in the full phase diagram because of the presence 
of glassy minima with lower free energy. At high polydispersity,
when crystallization does not occur, our calculations show that the glassy
state can be an equilibrium phase for this system. 

We are grateful to Surajit Sengupta and  Robin Speedy for useful discussions 
and Dhrubaditya Mitra for help in computation. PC would like to thank SERC (IISc) for providing
the necessary computation facilities and JNCASR for providing financial support.

\end{document}